\documentclass[pre,aps,superscriptaddress,showpacs,twocolumn]{revtex4}
\usepackage{epsfig}
\usepackage{latexsym}
\usepackage{graphicx}
\usepackage{amssymb}
 
\begin{document}
\title{Hyperdynamics for entropic systems: 
time-space compression and pair correlation function approximation}

\author{Xin Zhou \footnote{ Email: xzhou@lanl.gov} }
\affiliation{Earth and Environmental Science Division, Los Alamos
National Laboratory, Los Alamos, NM 87545}

\author{Yi Jiang}
\affiliation{Theoretical Division, Los Alamos National Laboratory, 
Los Alamos, NM 87545}

\author{Kurt Kremer}
\affiliation{ Max-Planck-Institut f\"{u}r Polymerforschung,
Ackermannweg 10, D-55128 Mainz, Germany}

\author{Hans Ziock} 
\affiliation{Earth and Environmental Science Division, Los Alamos
National Laboratory, Los Alamos, NM 87545}

\author{Steen Rasmussen}
\affiliation{Earth and Environmental Science Division, Los Alamos
National Laboratory, Los Alamos, NM 87545}

\date{\today}

\begin{abstract}

We develop a generalized hyperdynamics method, which 
is able to simulate slow dynamics in atomistic general 
(both energy and entropy-dominated) 
systems. We show that a few functionals of the pair correlation
function, involving two-body entropy, form a low-dimensional 
collective space, which is a good approximation that is able to 
distinguish stable   
and transitional conformations.  A bias potential, which raises the
energy in stable regions, 
is constructed on the fly. 
We examine the 
slowly nucleation processes of a
Lennard-Jones gas and show that our new method can generate 
correct long time dynamics without a prior knowledge. 
\end{abstract}


\maketitle

Molecular dynamics (MD) simulations are typically limited to a
time scale of less than a microsecond, so many interesting slow
processes in chemistry, physics, biology and materials science cannot
be simulated directly. 
Recently, new methods, including kinetic Monte 
Carlo, transition path ensemble methods, minimal action/time
methods and the constrained MD 
simulation~\cite{BortzKL1975,BolhuisCDG2002,OlenderE1996,LaioP2002}, 
have been developed
to study the slow processes (for a review see~\cite{Elber2005}).   
They all require a prior knowledge of the system, which is
often hard to obtain, and they can
only deal with a few special processes inside a small part of
the configurational space of the system. 
 
For many systems, the interesting dynamics are governed by the
infrequent, fast transitions between meta-stable regions; yet the systems
spend most of their time in the stable regions, whose dynamics can be well 
described by some time-averaged properties. 
Hence 
we would coarse grain the stable configurations while keeping the needed details
in the unstable regions that define the transitions.
Hyperdynamics, as developed by Voter~\cite{Voter1997} 
is an example of such a coarse-graining method. 
The hyperdynamics method treats the potential wells as 
the stable conformational regions that are separated by the saddle regions. 
A bias potential is designed to lift the energy of the system in these wells, 
while keeping the saddle regions intact. Dynamics on the
biased potential leads to accelerated 
evolution from one stable region to another. 
Based on transition state theory, 
the realistic escape time 
$t_{real}$ 
from the wells can be reproduced,  
$t_{real} = {\Delta t} \sum_{i} \exp [\beta \Delta V(r(t_{i}))]$,
where $\Delta t$ is the time step of MD, $\Delta V(r)$ is the applied bias 
potential along the simulated trajectory $r(t)$. 
This method has been applied successfully to systems in
which the relevant states correspond to deep wells in the potential energy, 
with dividing surfaces at the energy ridge tops separating these 
states~\cite{MironF2004}. 
It has however not been clear how to apply hyperdynamics in cases where the 
transitions are dominated by entropic considerations.
In such cases, the potential energy alone is not enough to distinguish  
stable and transition regions since some conformations with similar 
energy might belong to stable and transition regions, respectively. 
A complication that occurs even when trying to apply hyperdynamics to 
solids with fairly clearly defined stable regions is that after applying the 
bias potential, the energy landscape becomes much flatter and the system can 
start to have entropic-like characteristics. 
These effects limit the improvement in the simulation rate that can be 
achieved by the hyperdynamics method over the direct MD approach.
Thus, although there are some attempts in the
literature~\cite{RahmanT2002,HamelbergSM2005b} 
to apply 
hyperdynamics to enhance conformational sampling in bio-systems, generally, 
accurate slow dynamics or kinetic rates  
 can only be expected for relatively simple solids or low dimensional 
systems. 
 
In this Letter, we derive a more general hyperdynamics method that 
can be used to access longer time scales in fluids. We
present explicit conditions for applying this method. We use the
pair correlation function as a reliable means of identifying the 
important 
conformations and then construct the appropriate bias potentials in a
lower-dimensional  space while the simulation runs. We then examine 
the performance of this method by looking at 
the slow gas-liquid transition in a system of $N$ 
identical Lennard-Jones particles. 

We begin this process by introducing a time-compressing transformation, 
$d \tau = a(r) d t$, 
where $d \tau$ is a pseudo-time step, $d t$ is the real time step, and the 
local dimensionless compression factor is given by a conformational function 
$a(r)$, which is $\le 1 $.  
Thus the trajectory, $r(t)$, can be rewritten as 
$r(\tau) = r(\tau(t))$ in 
a shorter pseudo-time interval, 
$\tau = t \int d r \ D(r;r(t);t) \ a(r)$,
where $D(r;r(t);t)$ is the distribution of $r(t)$ in the interval 
$[0, t]$.
The compressed trajectory $r(\tau)$ satisfies a new equation of motion,
\begin{eqnarray} 
{d \over d \tau} {\mathbf P}_{i} = 
-\frac{\partial V}{\partial {\mathbf R}_{i}} 
+\sum_{j} \frac{{\mathbf P}_{i} }{M} {\mathbf P}_{j} \cdot 
\frac{\partial }{\partial {\mathbf R}_{j}} \ln a(r)
\label{newtonequation}
\end{eqnarray}
where ${\mathbf P}_{i}= M d {\mathbf R}_{i}/d \tau $, $M$ is the transformed 
mass, equal to $m a^{2}(r)$, and the $m$ is the real mass of the particles.
$V(r)$ is the 
potential energy, and $r$ is the simple donation of the position vectors 
${\mathbf R}_{i}$ ($i=1,\cdot \cdot \cdot N$) of all the $N$ particles.  
At first glance it would not appear to be 
advantageous to directly generate 
$r(\tau)$ from Eq.(\ref{newtonequation}) as  
very short time steps are necessary due to the small values of $M$.
However, if we only focus on the long-time dynamics, 
we can use a smoother pseudo-trajectory 
${\cal R}(\tau)$ to replace $r(\tau)$, provided that one requires the 
reproduced time from ${\cal R}(\tau)$ to be the same as that from $r(\tau)$.
Thus, a  
sufficient condition to replace $r(\tau)$ with ${\cal R}(\tau)$ is that their 
distributions are the same. 

In general, the distribution along a finite-length trajectory is not easily 
known. 
However, in the time-consuming regions (stable regions), similar conformations 
would be visited many times 
even during a finite simulation time. Thus we can assume the 
distribution can be approximated by $D(r;r(t);t) \propto \exp(-\beta V(r))$. 
Many methods might be used to generate ${\cal R}(\tau)$ with the required 
distribution. 
One simple method is to use a realistic trajectory corresponding to the local 
equilibrium of a 
new potential $U(r) = V(r) - k_{B} T \ln a(r)$. 
Actually, based on the same local equilibrium supposition, if one replaces   
the kinetic energy term of Eq.(\ref{newtonequation}) by  
its ensemble averaged value,
$<{\mathbf P}_{i} {\mathbf P}_{j}/M> = k_{B} T \delta_{ij}$, 
Eq.(\ref{newtonequation}) is indeed the equation of motion of the particles 
with smaller mass $M$ under the new 
potential $U(r)$.  
Outside of the time-consuming regions, we choose $a(r)$ as unity so that 
the MD time is realistic.  
If we select small $a(r)$ only in the potential well regions, we effectively have 
the 
hyperdynamics presented by Voter~\cite{Voter1997}. 
The condition for the use of compressed time is that the simulated trajectory under 
$U(r)$ has the equilibrium distribution in all the biased (time-compressed) 
regions. Simply, we can suppose that the distribution is locally given by the 
Boltzmann one in the 
conformational regions where the value of the distribution is larger than a 
critical value. In comparison with the transition state theory in original 
hyperdynamics implementation, the new approach makes it is easier to both 
determine the proper regions to bias and to design the appropriate 
bias potential. 
 
In solids, as the number of conformations inside stable regions (potential 
energy wells) is often small and the distribution of long-time 
trajectories can reach local equilibrium, 
the hyperdynamics method works very well. 
However, in entropy-dominated systems (e.g. gas and liquids), 
the number of conformations (entropy) inside time-consuming regions may be 
huge, and thus the Boltzmann distribution may not be reached in finite MD time. 
In this case, 
we average  the neighboring distribution of the trajectory,
${\bar D}(r;\sigma) = \int_{\sigma} D(r) dr$, where 
$\sigma$ is the size of selected neighbors. 
For the smoothed distribution, in time-consuming regions, we can 
expect,
${\bar D}(r;\sigma) \propto \int_{\sigma} \exp(-\beta V(r)) d r$.
Thus, by selecting smooth $a(r)$ functions, the realistic time propagator 
can still be reproduced from the physical trajectory of $U(r)$, provided  
the averaged distribution of the $U(r)$ trajectory satisfies the equation. 
 
As an illustration, we compress the $3N$-dimensional flat conformational space $r$ 
to a curved $3N$-dimension $q$ space,
	$d q = {\mathbf A}(r) d r$,
where ${\mathbf A}(r) = \frac{\partial q}{\partial r}$ 
is the Jacobian of the transformation.
In $q$ space, the trajectory $q(t)=q(r(t))$ satisfies a new equation of 
motion with 
a positive symmetric mass matrix 
${\mathbf M} = {\mathbf B}^{T}(r) \ {\mathbf m} \ {\mathbf B}(r)$, where 
${\mathbf B}^{T}$ denotes the transpose, ${\mathbf m}$ is the real mass 
diagonal matrix, and
${\mathbf B} = {\mathbf A}^{-1}$. 
 Not losing any generality, we choose a diagonal ${\mathbf M}$ by 
rotating the $d q$ space, and the new equation of motion is, 
 \begin{eqnarray} 
{d \over d t} {\mathbf P} = -\frac{\partial V}{\partial q} 
+ \sum_{j} \frac{ P^{2}_{j} }{ 2 { \mathbf M }_{j} } \ 
\frac{\partial \ln M_{j}(q) } {\partial q}, 
\label{newton3}
\end{eqnarray}
where ${\mathbf P}$ and $P_{j}$ are the generalized momentum and its $j$ 
component, respectively, and 
 $M_{j}$ is the $j$ diagonal element of ${\mathbf M}$.
If one replaces $\frac{P^{2}_{j}}{M_{j}}$ with $k_{B} T$, 
Eq.(\ref{newton3}) is the equation of motion of heavier particles 
(mass $M_{j}$) under a new potential,
$W(q) = V(q) + k_{B} T \log J(q)$, where $J(q) = det({\mathbf A})$ is the 
determinant of the Jacobian ${\mathbf A}$ of the transformation. 
Here, $W(q)$ is indeed the free energy of system in $q$ space. 
 Thus, by inhomogeneously compressing the conformational space, we transform 
the original 
entropy-dominated $r$ space to a energy-dominated $q$ space with 
effective potential $W(q)$. 
If the trajectory $q(t)$ of $W(q)$ corresponds to the Boltzmann distribution 
in some time-consuming regions, the corresponding trajectory $r(q(t))$ 
represents a local equilibrium in $r$ space. 
Thus we can bias the effective potential $W(q)$ in $q$ space to extend 
the MD time scale. Actually, this can be done directly in $r$ 
space without using the explicit transformation ${\mathbf A}(r)$. 
 
The keys to successfully apply hyperdynamics are distinguishing the 
conformations and designing suitable bias potentials 
 $\Delta V(r)$ for the entire conformation space $r$.   
Obviously, $\Delta V(r)$ 
should have the same symmetry as $V(r)$. 
 Considering a simple case of $N$ identical particles, 
 we rewrite the conformational vector $\{ {\mathbf R}_{j} \}$ 
($j=1, \cdot \cdot \cdot, N$) as a density field,
${\hat \rho} ({\mathbf x}) = \sum_{j} \delta({\mathbf x}-{\mathbf R}_{j})$.
Here both ${\mathbf x}$ and ${\mathbf R}$ are the normal $3$-dimension spatial 
vectors.
 Since the neighboring conformations are identical in the viewpoint of slow dynamics,  
$\hat \rho(\mathbf x)$ can be averaged to get a smooth function,
$\bar \rho(\mathbf x)$, by for example using a  
Gaussian function 
to replace the Dirac-$\delta$ 
function. 
 If the width of the Gaussian function is small, 
$\bar \rho(\mathbf x)$ can be used to identify different conformations.
 Here we used a functional space to replace the $3N$-dimension conformation space, 
but actually, the physically allowed $\bar \rho(\mathbf x)$ only occupies 
a very small part of the functional space. 
 By neglecting multi-body correlations and directional correlations, 
we can approximate the density field $\bar \rho(\mathbf x)$ (or 
conformations) by using
some bin-averaged values of the radial pair correlation function $g(x)$ of the 
conformations (here, $x$ donates the length of the spatial 
vector ${\mathbf x}$),   
$g_{i} = {1 \over \Delta} \int_{\Delta} g(x + x_{i}) d x$.
Here $g_{i}$ is an average value of $g(x)$ in a small bin  
 $(x_{i} - \Delta/2, x_{i} + \Delta/2)$. 
 Thus, each conformation corresponds to the group of $g_{i}$ that defines a 
point in $g$ space. The spatial neighbors of the conformation and their 
symmetric companions will also correspond to the same $g$ point.  
If the bin size is very small, all conformations with the same $\{g_{i}\}$ are 
identical in the slow dynamics viewpoint and thus the bias potential of 
hyperdynamics can be written as a function in the low-dimension $g$ space, 
$\Delta V({\mathbf R}^{N}) = f(\{ g_{i}({\mathbf R}^{N})\})$. 
To better identify conformations even when larger bin sizes are used,  
some important dynamics-related physical variables, such as the  
potential energy $V({\mathbf R}^{N})$, 
can be added into the $\{ g_{i} \}$ variable group. 
Another important variable is the two-body entropy, 
$S_{2} = - 2 \pi \rho \int [g(x) \ln g(x) - g(x) + 1] x^2 dx$.
which forms the main part (about $90$ percent) of the macroscopic 
excess entropy
~\cite{Green1952}. 
Similarly, it is also possible to use some other functional of $g(x)$ to 
replace some $g_{i}$. 
In special systems, it may be useful to add some 
special order parameters $O_{j}$  
to take into account possible multi-body correlations and thereby decrease 
the needed number of $g_{i}$. 
Finally, we have a group (of order $10$) of general collective variables 
denoted as $S=\{ S^{j}\}$, which might involve $V$, $S_{2}$, some $\{ g_{i} \}$ 
and some possible $\{ O_{j} \}$,  
to identify conformations and form an appropriate 
bias potential. 
In general, we construct the bias potential as 
$\Delta V(S({\mathbf R}^{N})) = k_{B} T f_{+}(\ln D(S)/D_{c})$,  
where, $D(S)$ is the distribution of the simulated 
trajectory, $D_{c}$ is the selected critical value. 
The function $f_{+}(z) = z$ for larger positive $z$, and smoothly approaches 
$0$ as $z$ decreases to $0$. 
The designed bias potential will generate a flatter
distribution in $S$ space while the biased regions are  
still visited often enough for the system to reach local equilibrium.  
The bias potential can be formed gradually. First, we generate a long 
non-biased MD trajectory to calculate $D(S)$ and form a (small) bias 
in some $S$ regions. Next a long 
trajectory is simulated in the biased system, which in turn generates the 
basis for the next 
bias. This process is repeated until the desired long realistic 
time is reached. Thus, we can gradually study dynamics at ever longer  
time scales, but at the expense of the details seen in fast dynamics. 
 With the biased potential, the bias force on particles can be calculated 
by the chain rule of differentiation, 
$\Delta {\mathbf f}_{i} = - \sum_{j} \frac{\partial \Delta V}{\partial S^{j}} 
\frac{\partial S^{j}}{\partial {\mathbf R}_{i}}$.
 
We have examined the general hyperdynamics method in a simple system of $N$ 
identical 
Lennard-Jones (LJ) particles and studied the slow 
gas/liquid transition in the $NVT$ ensemble. 
 We used 
the truncated and shifted LJ potential with $r_{c} = 2.5$, and the reduced 
units:
$\epsilon_{lj}=1$, $\sigma_{lj} =1$ and the mass of particles $m=1$. 
 The temperature is fixed as $T = 0.613$. 
The velocity Verlet algorithm was used to integrate the Langevin equation 
of motion.
 Obviously, in such a system, the potential energy of transitional 
conformations ( liquid drops with critical size) is 
lower than that of the stable gas phase.  
Thus, some simple bias methods, for example, just lifting the energy of all 
lower-energy conformations~\cite{SteinerGW1998}, cannot work at all in such 
a system. Actually, for general entropy-important systems, using the potential 
alone is not enough to identify the transitional conformations and then form 
the bias potential. Only in some special low-dimension 
systems~\cite{HamelbergSM2005b} where the entropy is occasionally not 
important, exceptions may be found.

We initially studied a gas phase with a relatively large saturation ( a number 
density $n=0.02$ for a system of $N=1000$ particles) for which a    
liquid drop forms in normal MD without any bias. 
The phase transition happens in a narrow time window where the  
potential $V$, the two-body entropy $S_{2}$ 
and the pair correlation function $g(x)$ are found to change drastically 
in the gas/liquid phase transition process. 
However, inside each small $S_{2}$ range, we found $g(x)$ only 
fluctuates slightly around its average value. 
This shows that the two-body entropy $S_{2}$ integrates the information of 
the $g(x)$.  
Thus, in this simple system, we use only two functionals of $g(x)$, namely 
$V$ and $S_{2}$, to form the collective space while designing the bias potential. 
By examining the $g_{i}$, which are available from our simulations, we were 
able to show that $V$ and $S_{2}$ can sufficiently identify 
transition states and stable states, at least in this simple system. 
Since $S_{2}$ gives the main part of the entropy, we expect that $S_{2}$ and 
$V$ are also leading collective variables even in more complex systems. 

\begin{figure}
\includegraphics[width=8cm]{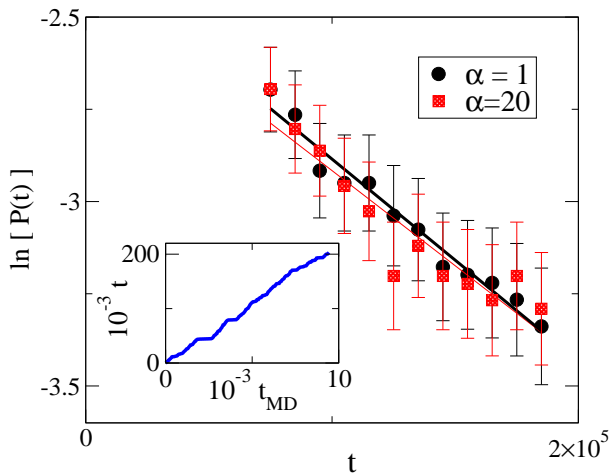}
\caption{(Color online). The distribution of transition time of the gas phase in 
direct MD ($\alpha=1$) and biased MD ($\alpha = 20$) simulations. The inverse 
slope of the least-squares fit to the points (solid line) gives the lifetime 
of the gas phase. $n = 0.014$, $N=400$ and $T=0.613$. Inset: the time boost 
in a typical biased MD simulation is shown. 
}
\label{fig1}
\end{figure}

\begin{figure}
\includegraphics[width=8cm]{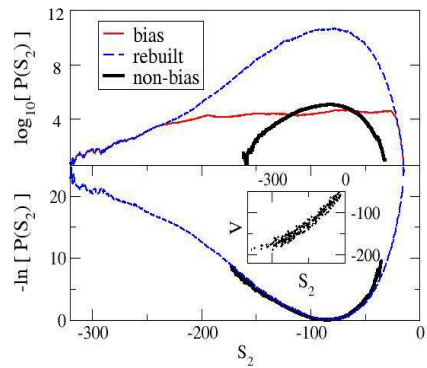}
\caption{(Color online). Top: the distributions of two body entropy $S_{2}$ from 
non-biased and biased simulations. The rebuilt distribution of the bias 
simulation is also shown. Here, $n = 0.008$, $N=1000$ and $T=0.613$. 
Bottom: the free energy profiles from the non-biased and biased simulations 
are compared. The inset shows the simulated  
samples in the $(S_{2},V)$ space. The observed liquid phase 
does not show here.
}
\label{fig2}
\end{figure}

When we decrease the density (or saturation) of the LJ system, the 
lifetime ($\tau = 1/k$, $k$ is the transition rate) of 
the gas phase increases drastically. For example,  
$\tau \sim 10^{4}$ when the density $n=0.016$, but increases by about a factor 
of $20$ when the density is only slightly decreased to $n = 0.014$. Even in 
this density region, 
direct simulation of the gas/liquid transition is still possible. 
Thus, we compared the transition kinetics with and without the use of a bias. 
Fig.(\ref{fig1}) shows that the distribution of the gas to liquid transition 
waiting time $t$ is exponential in $t$ from both non-biased and 
biased simulations, 
$\ln P(t) \propto - t/\tau$. 
The boost factor $\alpha$ resulting from the hyperdynamics method, which 
characterizes the 
average gain in the rate at which time advances relative to direct MD, 
is about $21$, as shown in the inset of Fig.(\ref{fig1}) ($\alpha = 1$ for 
unbiased MD simulation).
We also directly calculate $\tau$ by averaging the transition time of full
simulations for a better comparison.
We found that in the biased MD case, $\tau = 2.02 \times 10^{5}$, 
in excellent agreement with the direct MD result, $\tau = 2.07 \times 10^{5}$. 
In our current simulations, we gradually increase the bias potential until our 
desired transition can happen in the usual MD steps. 
Thus we need and can apply larger bias potentials in the 
lower-density LJ systems.
 At $n=0.012$, the phase transition is detected  
while $\alpha$ is of order of $100$. At still lower density, for example,  
$n=0.01$, it is very difficult (if not impossible) to observe the transition 
using direct MD simulations. However, with our method 
we can still easily detect the gas/liquid transition. 
 Fig.(\ref{fig2}) shows the results for $n=0.008$ and $N=1000$. 
The distribution $P(S_{2})$ 
 is flatter and broader in the biased simulation, indicating that the system 
 visits a larger conformational space.  
The lower  
panel of Fig.(\ref{fig2}) shows the reproduced  
free energy profiles from the distribution of the biased simulation 
($\alpha \approx 10^{6}$). It greatly agrees with that from a non-biased 
simulation in the region where the direct MD is possible. The inset of 
Fig.(\ref{fig2}) shows the distribution of samples in the $(V,S_{2})$ space. 
The shown dense region ($S_{2}>-300$) which corresponding to the gas 
phase is biased due to its higher distributed density of samples. The 
lower-density region shown in the inset ($S_{2}<-300$) corresponds to the 
transition region ( the liquid phase which $S_{2}$ is far smaller $-300$ does 
not shown) was not biased. From the obtained distribution, we know $S_{2}$ is 
actually a good reaction coordinate, the difference between the rebuilt and 
biased $\log_{10} P(S_{2})$, shown in the higher panel of 
Fig.(\ref{fig2}), corresponds the profile of the applied bias potential 
(with a factor $k_{B} T \ln 10$). 

To summarize, 
we have expanded the hyperdynamics method to more general cases by 
inhomogeneously compressing time and conformational space. Our approach directly 
generates an explicit general method to design the bias potential.
In simple systems, 
a few functionals of the pair correlation function provide a good 
approximation of the density field for identifying the important 
conformations and for constructing the bias potential without prior knowledge 
of the conformational space. 
The method is expected to be applicable in more complex fluids where even more 
collective variables might be needed. 
 
This work was supported by the US DOE under contract W-7405-ENG-36 
with LDRD No. LA-UR-06-2794. 
We are grateful to A. F. Voter, H. Chen for 
stimulating discussions, comments and suggestions. 
 
\newpage

\newpage

\end{document}